\documentclass[aps,prb,twocolumn,superscriptaddress,amsmath,amssymb,floatfix,longbibliography]{revtex4-2}

\usepackage{graphicx}
\usepackage{bm}
\usepackage{physics}
\usepackage{mathtools}
\usepackage{color}
\usepackage{hyperref}
\usepackage{comment}

\newcommand{\runinhead}[1]{\textit{#1}.--}

\begin{document}

\title{Strain-Induced Helical Superconductivity and the Zero-Field Diode Effect}


\author{Raigo Nagashima}
\affiliation{Institute for Theory of Condensed Matter, Karlsruhe Institute of Technology, Karlsruhe, 76131, Germany}
\author{J\"org Schmalian}
\affiliation{Institute for Theory of Condensed Matter, Karlsruhe Institute of Technology, Karlsruhe, 76131, Germany}
\affiliation{Institute for Quantum Materials and Technologies, Karlsruhe Institute of Technology, Karlsruhe, 76131, Germany}


\date{\today}


\begin{abstract}
Motivated by the strain-induced zero-field superconducting diode effect observed in PbTaSe$_2$, we identify a mechanism by which strain generates nonreciprocal superconducting transport without magnetism or an external magnetic field. Uniaxial strain mixes a dominant $s$-wave order parameter with a subdominant two-component pairing channel and enhances their symmetry-allowed Lifshitz coupling. Beyond a critical strain, this coupling drives a transition into a helical state with spontaneously selected finite Cooper-pair momentum and broken time-reversal symmetry. The resulting diode effect is generically non-monotonic in strain and exhibits distinct responses for currents parallel and perpendicular to the residual mirror plane, consistent with experiment. Our theory predicts that reversing the principal strain switches the diode direction by $90^\circ$, while shear strain rotates it continuously. These results establish strain as a symmetry-selective means of creating, controlling, and diagnosing spontaneous helical superconductivity.
\end{abstract}

\maketitle

\runinhead{Introduction}
The superconducting diode effect (SDE) is a non-reciprocal phenomenon in which the magnitude of the critical current of a superconductor depends on the direction of the applied current. This transforms a superconductor into a rectifier—a non-dissipative analog of a semiconductor diode—while maintaining zero resistance for current flow. Following important early experimental studies~\cite{Swartz1967, Cerbu2013} and theoretical proposals~\cite{Levitov1985, Geshkenbelh1986, Edelstein1996, Vodolazov2005}, interest in the SDE has grown rapidly, driven by the observation of the effect in an artificial superlattice~\cite{Ando2020} and its subsequent discovery in a variety of bulk materials~\cite{Lyu2021, Bauriedl2022, Strambini2022, Hou2023, Kealhofer2023}, as well as in Josephson junctions~\cite{Pal2022, Turini2022}.
\begin{figure}
    \centering
    \includegraphics[width=1.0\linewidth]{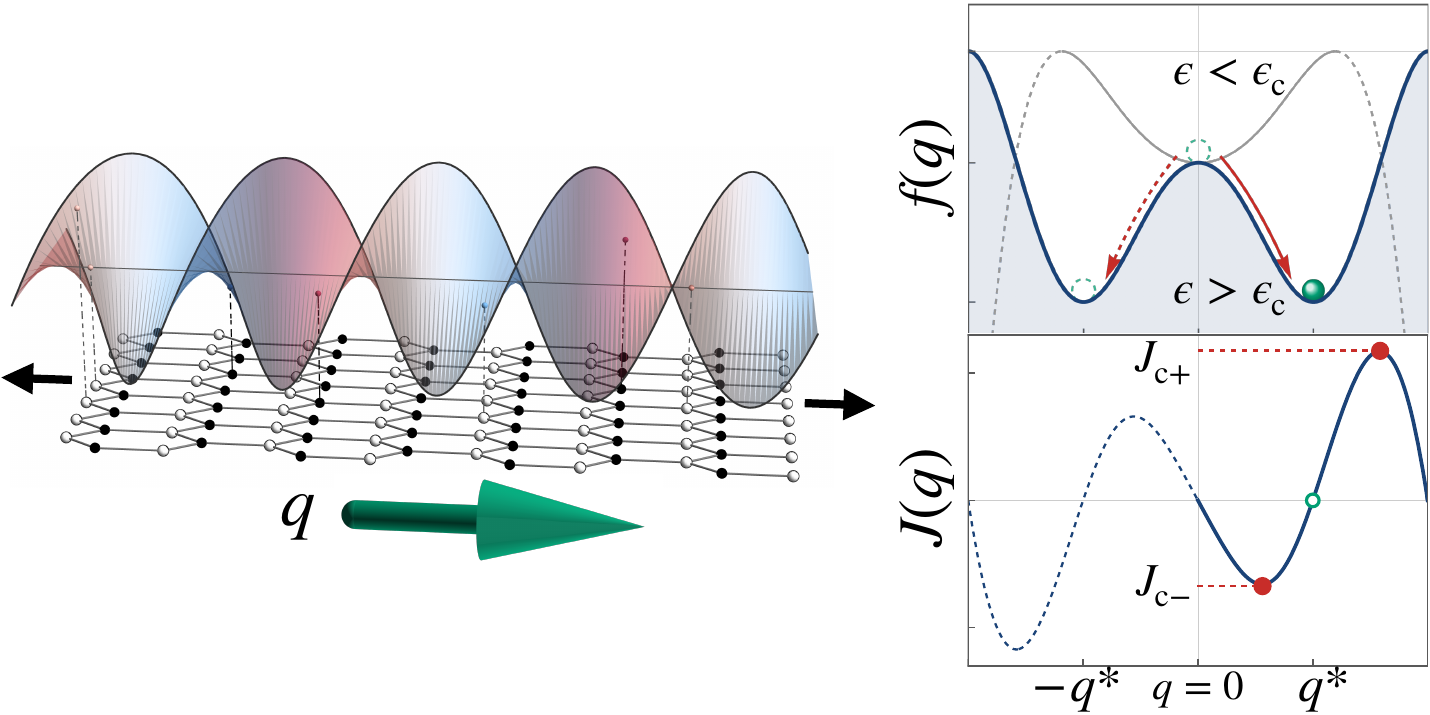}
    \caption{
    (Left) Schematic of a strain-induced spontaneous helical state. The color gradient represents the spatially varying phase of the superconducting order parameter. (Right) Schematic free energy $f(\bm{q})$ (top) and corresponding current density $J(\bm{q})$ (bottom). When the strain $\epsilon$ exceeds the critical value $\epsilon_{\mathrm{c}}$, the system transitions to a helical state with finite momentum $\bm{q}^*$. The superconducting diode effect is quantified by the difference between the magnitudes of the positive and negative critical currents, $J_{\mathrm{c}+}$ and $J_{\mathrm{c}-}$, respectively.
    }
    \label{Schematic}
\end{figure}

Broken time-reversal symmetry (TRS) and inversion symmetry are generally regarded as necessary prerequisites for the SDE~\cite{Nadeem2023, Moll2023, Nagaosa2024, Ma2025, Shaffer2025}.
In most experiments, the effect is induced by an applied magnetic field that explicitly breaks TRS.
Field-free SDEs have also been reported. Some rely on ferromagnets~\cite{Narita2022, Jeon2022} or other forms of spontaneous time-reversal symmetry breaking (TRSB), which likely originate either in the normal state~\cite{Wu2022, Lin2022, Wan2024, Le2024, Ge2026, Nagata2025} or in an unconventional pairing state formed at a tunneling junction~\cite{Zhao2023, Ghosh2024, Qi2025}.
Current-induced vortices have also been proposed as a mechanism underlying the diode effect~\cite{Lee1999, Villegas2003}.

The most widely discussed intrinsic mechanism for the SDE in the presence of a magnetic field~\cite{Daido2022, Yuan2022, He2022} is based on the interplay between Rashba spin-orbit coupling and the Zeeman effect, which gives rise to a helical superconducting state with a finite Cooper-pair momentum $\bm{q}$~\cite{Agterberg2003, Smidman2017}.
In this state, the superconducting order parameter acquires the spatial modulation $e^{\mathrm{i}\bm{q}\cdot\bm{r}}$.
This mechanism predicts the depairing current~\cite{Tinkham}, the maximum possible critical current, and therefore the intrinsic limit of the SDE. Considerable theoretical effort has been devoted to understanding the origin of the SDE from various perspectives~\cite{Ilic2022, Legg2022, Zhang2022, Scammell2022, Daido2022_2, Chazono2023, Banerjee2024, Nunchot2024, Nakamura2024, Shaffer2024, Banerjee2024_2, Daido2025, Bankier2025, Hasan2025, Shaffer2025, Chen2025, Zhuang2026, Nunchot2026}.

Recently, a zero-field SDE induced by uniaxial tensile strain, without explicitly breaking TRS, was reported in $\textrm{PbTaSe}_{2}$~\cite{Liu2024}. This material has trigonal symmetry with point group $D_{3h}$ and is therefore non-centrosymmetric. Applying uniaxial strain along the armchair direction lowers the symmetry to $C_{2v}$ while preserving a mirror plane. Under strain, the zero-field SDE appears only for currents applied parallel to the mirror plane and is absent for currents flowing perpendicular to it. Moreover, upon applying an out-of-plane magnetic field $B_{z}$, the SDE measured parallel to the strain axis exhibits an even dependence on $B_z$, whereas the conventional SDE, measured perpendicular to the strain axis, displays the usual odd dependence on $B_{z}$. Similar behavior has also been observed in NbSe$_{2}$~\cite{Li2025}.

In this work, we show that strain can induce spontaneous TRSB by driving a transition from a conventional $s$-wave superconducting state to a translation-symmetry-breaking helical state. This transition is enabled by the strain-induced enhancement of a symmetry-allowed linear-gradient coupling (Lifshitz invariant~\cite{Kamatani2022, Nagashima2024, Nagashima2025}) between the $s$-wave order parameter and a subdominant, two-component unconventional pairing state.
Once the strain exceeds a critical threshold, the system enters a helical state with nonzero Cooper-pair momentum, giving rise to a zero-field SDE (see Fig.~\ref{Schematic}).
We further investigate the dependence of the SDE on an out-of-plane magnetic field and predict its response to additional shear strain, providing an experimentally testable means of falsifying our theory.

\runinhead{Ginzburg-Landau free energy under strain}
Suppose the primary superconducting order parameter $\psi_{1}$ in the absence of strain belongs to the trivial irreducible representation (irrep) $A_{1}'$ of $D_{3h}$, consistent with a fully gapped state seen in experiments~\cite{Ali2014, Pang2016, Sankar2017, Wilson2017}, and is described by the  usual Ginzburg-Landau expansion
\begin{equation}
f_1= a_{1}|\psi_{1}|^{2} + \frac{b_1}{2}|\psi_{1}|^{4} + c_{1}|\bm{D}\psi_{1}|^{2}.
\end{equation}
Here, $a_{1} = a_{1,0}(T-T_{\text{c}1})$ with $a_{1,0}$, $b_{1}$, $c_{1}$, positive constants, and $\bm{D} = \nabla - \mathrm{i}e^{*}\bm{A}/\hbar$.
We do not distinguish between different coefficients $c_{1}$ for in-plane and out-of-plane gradients, an assumption that can easily be lifted.

In the experiment of Ref.~\cite{Liu2024} uniaxial strain which contributes to $\epsilon_{x^{2} - y^{2}}$ was applied, where $\left(\epsilon_{x^{2} - y^{2}},-2\epsilon_{xy}\right)$ forms a doublet that transforms under the $E'$ irrep of  $D_{3h}$. Strain  mixes $\psi_{1}$ with  a secondary superconducting order parameter $\psi_{2}$ that also belongs  to $E'$:
\begin{eqnarray}
    f_{\epsilon}=\eta\Big( \epsilon_{x^{2} - y^{2}}\psi_{1}^{*}\psi_{2x} - 2\epsilon_{xy}\psi_{1}^{*}\psi_{2y} + \textrm{c.c.} \Big).
\end{eqnarray}
This mixing merely reflects that $\psi_{2}$, which transforms trivially under the remaining mirror operation, becomes part of the trivial representation of the symmetry group $C_{2v}$ of the strained crystal.
The secondary order-parameter is governed by its own free-energy expansion $f_2= \sum_{\alpha=x,y}\left(a_{2}|\psi_{2\alpha}|^{2}  + c_{2}|\bm{D}\psi_{2\alpha}|^{2} \right)$. We use $a_2>0$ to reflect the fact that $\psi_{2}$ would not, by itself, condense.
Accordingly, we neglect non-linear, quartic terms in $\psi_{2}$.
Additional symmetry-allowed gradient terms do not change the conclusion of our analysis.

The strain-induced channel mixing $f_\epsilon$, while important for our theory, does not suffice to explain the SDE. This is achieved by including another symmetry-allowed coupling of the two order parameters through  Lifshitz terms
\begin{eqnarray}
 f_{\rm L} &=&  \lambda\Big( \psi_{1}^{*}D_{x}\psi_{2x} + \psi_{1}^{*}D_{y}\psi_{2y} + \textrm{c.c.}  \Big) \notag \\
&+& \mu\epsilon_{x^{2} - y^{2}}\Big(\psi_{1}^{*}D_{x}\psi_{2x} - \psi_{1}^{*}D_{y}\psi_{2y} + \textrm{c.c.}  \Big) \notag \\
&+& 2\mu\epsilon_{xy} \Big(\psi_{1}^{*}D_{x}\psi_{2y} + \psi_{1}^{*}D_{y}\psi_{2x} + \textrm{c.c.}  \Big),   
\end{eqnarray}
where we included linear gradient terms in the absence and presence of strain, respectively. The coefficients $\eta$, $\lambda$, and $\mu$ are real for a time-reversal symmetric normal state.

The Ginzburg-Landau free energy of the two superconducting order parameters and up to first order in strain is $f=f_1+f_2+ f_{\epsilon} +f_{\rm L}$.
We will show that it yields, under specific conditions, a helical TRSB pairing state that gives rise to a zero-field SDE.
To be consistent with the experiment, we first assume $\epsilon_{xy}=0$ but $\epsilon_{x^{2}-y^{2}}\neq 0$.
Then, only $\psi_{2x}$ will be induced by strain (at least as long as $\lambda/(\mu \epsilon_{x^2-y^2}) >0$, see below).
We fix $\bm{A}=\bm{0}$, focus on $T<T_{\text{c}1}$, and denote $\epsilon_{x^{2}-y^{2}}$ as $\epsilon$ for the moment.
The more general case involving shear strain or a magnetic field is discussed below.

\runinhead{Spontaneous helical state and zero-field SDE}
We solve the Ginzburg-Landau theory with the ansatz 
\begin{equation}
    \psi_{i} = \psi_{i,0} + \psi_{i,+}e^{+\mathrm{i}\bm{q}\cdot\bm{r}} + \psi_{i,-}e^{-\mathrm{i}\bm{q}\cdot\bm{r}} \ (i=1,2x).
\end{equation}
For sufficiently small strain and Lifshitz coupling, the stable solution is the homogeneous one with $\psi_{i,0}\neq0$ and $\psi_{i,\pm }=0$.
A helical state is realized when the free energy is minimized under the condition that $\psi_{i,0}=0$ and either $\psi_{i,+}$ or $\psi_{i,-}$, but not both, are nonzero. 
Considering finite-momentum solutions $\psi_{i,\pm}\neq 0$ yields
\begin{equation}
    \psi_{2x\pm}=\frac{\eta\epsilon\pm \mathrm{i}\left(\lambda+\mu\epsilon\right)q_{x}}{a_{2}+c_{2}q^{2}}\psi_{1}.
\end{equation}
We  insert this into $f$ and obtain the effective theory of the primary order parameter
\begin{align}
    f &=\left|a_{1}\right| R(\bm{q})(|\psi_{1,+}|^{2} + |\psi_{1,-}|^{2}) \notag \\
    &+ \frac{b}{2}(|\psi_{1,+}|^{4} + |\psi_{1,-}|^{4} + 4|\psi_{1,+}|^{2}|\psi_{1,-}|^{2}),
\end{align}
with a dimensionless coefficient that contains the momentum dependence
\begin{equation}
R(\bm{q}) = -1+\xi_{1}^{2}q^{2} - \gamma\frac{1+\ell^{2}q_{x}^{2}}{1+\xi_{2}^{2}q^{2}},
\end{equation}
Here $\xi_{i}=\sqrt{c_{i}/\left|a_{i}\right|}$ are the correlation lengths of the two uncoupled order parameters, where the primary order parameter is expected to have longer-ranged correlations: $\xi_1 > \xi_2$.
In addition, $\ell=\left|\frac{\lambda+\mu\epsilon}{\eta\epsilon}\right|$ is a length scale associated with the strain and Lifshitz couplings; at small strain we expect $\ell\gg \xi_2$.
Finally, $\gamma=\frac{(\eta\epsilon)^{2}}{\left|a_{1}\right|a_{2}}$ is a dimensionless coupling constant.
While $\gamma\propto \epsilon^2$ implies a vanishing coupling at zero strain, the denominator vanishes as $\left|T-T_{\text{c}1}\right|$.

There are two sets of stationary points: $(\psi_{1,+},\psi_{1,-})=e^{\mathrm{i}\theta}(\sqrt{a_1 R/b},0)$ and $e^{\mathrm{i}\theta}(0,\sqrt{a_1 R/b})$, as well as $(\psi_{1,+},\psi_{1,-})=e^{\mathrm{i}\theta}\sqrt{a_1 R/3b}(1,1)$.
The free energy of the former is $f=-a_1^2R^{2}/2b$, while the latter $f=-a_1^2R^{2}/3b$.
Hence, the TRSB helical state is  also lower in energy than the homogeneous solution if
\begin{equation}
    \frac{\gamma(\ell^{2} - \xi_{2}^{2}) }{\xi_{1}^{2}}>1,
    \label{qx_condition}
\end{equation}
i.e., when the strain and Lifshitz couplings exceed threshold values.
The ground-state momentum $\bm{q}=(q_{x}^{*}, 0)$ is  then given by
\begin{equation}
    q_{x}^{*} = \pm \frac{1}{\xi_{2}}\sqrt{ \sqrt{ \frac{\gamma(\ell^{2} - \xi_{2}^{2})}{\xi_{1}^{2}}} - 1}.
\end{equation}
One of these solutions is spontaneously chosen, thereby breaking TRS. We can then determine the current $\boldsymbol{J}=-\frac{\delta f}{\delta\boldsymbol{A}}=\frac{e^{*}}{\hbar}\frac{\delta f}{\delta\boldsymbol{q}}$, see Fig.~\ref{Schematic}. Hence,  non-reciprocity arises when the system occupies one of the two degenerate minima at $\pm q^*$, with the reverse critical current $J_{c-}$ determined by the local extremum of $J(q)$ between the occupied minimum and the intervening barrier.

The condition, Eq.~(\ref{qx_condition}) for the transition, expressed in terms of the parameters of the theory reads $\lambda^{2}+2\mu\lambda\epsilon+\left(\mu^{2}-\frac{c_{2}}{a_{2}}\eta^{2}\right)\epsilon^{2}>a_{2}c_{1}$. 
Since the dominant temperature dependence of the primary order parameter $a_{1}(T)$ does not enter, the criterion for TRSB is at most weakly temperature-dependent. We also stress that it is in principle possible to fulfill this condition already at zero strain, if $\lambda^{2}>a_{2}c_{1}$. This does not seem to be the case for $\textrm{PbTaSe}_{2}$, but could be obeyed in different materials.
Finally, we note that the modified coefficient $|a_{1}|R(\bm{q}^{*})$ from $a_{1}$ leads to an increase in the transition temperature
\begin{equation}
   T_{\text{c}}(\epsilon) =T_{\text{c}1}+ \frac{(\eta\epsilon)^{2}}{a_{1,0}a_{2}} + \frac{c_{1}}{a_{1,0}}\xi_{2}^{2}(q_{x}^{*})^{4}
\end{equation}
with strain, which is further enhanced upon entering the helical state with $q_x^*\neq 0$.

\begin{figure}[h]
    \centering
    \includegraphics[width=0.7\linewidth]{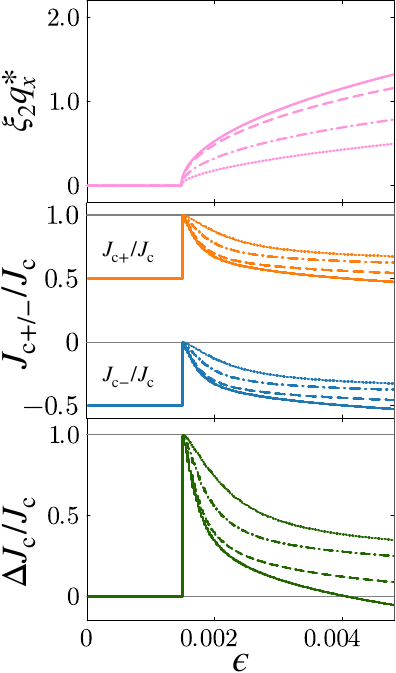}
    \caption{
    The strain dependence of $\xi_{2}q_{x}^{*}$ (above), normalized critical currents $J_{\mathrm{c}+/-}/J_{\mathrm{c}}$ (middle), and the normalized zero-field SDE (diode efficiency) $\Delta J_{\text{c}}/J_{\text{c}}$ (below) with $J_{\mathrm{c}}:=|J_{\mathrm{c}+}|+|J_{\mathrm{c}-}|$.
    We fix $\xi_{1}/\xi_{2}=5.48$, $\lambda/\eta=0.001\times\xi_{2}$, $\mu/\eta=3.73\times\xi_{2}$ and $\lambda^{2}/(c_{2}|a_{1}|)=1$, and show results for different $d/(c_{1}^{2}\xi_1^2)$:
$0.00055$ (solid), $0.0055$ (dashed), $0.055$ (dash-dotted), and $0.277$
(dotted).
    }
    \label{Fig_Strain_dep_q_J_SDE}
\end{figure}

Fig.~\ref{Fig_Strain_dep_q_J_SDE} shows the results for the critical current as a function of strain, where $\boldsymbol{J}_{\text{c}+/-}$ are defined in Fig.~\ref{Schematic}. Fig.~\ref{Fig_Strain_dep_q_J_SDE} shows the strain dependence of $\xi_{2}q_{x}^{*}$ (above), normalized maximum and minimum critical currents $J_{\text{c}+/-}/J_{\text{c}}$ (middle), and their amplitude difference (diode efficiency) $\Delta J_{\text{c}}/J_{\text{c}}$ (below) with $J_{\text{c}}=|J_{\text{c}+}|+|J_{\text{c}-}|$.
We find, at the onset of TRSB, perfect diode efficiency, similar to other approaches with continuous symmetry breaking~\cite{Yuan2022, Shaffer2024, Chakraborty2025}. See however our discussion below for caveats to these findings.
The SDE eventually decreases for increasing strain.
These results can also be understood analytically.
For $\xi_{2}q_{x}^{*}\ll1$, i.e., at the onset of TRSB, $|\boldsymbol{J}_{\text{c}+}|$ is of the order of the critical current in the absence of strain and TRSB, while $\boldsymbol{J}_{\text{c}-}\sim -\xi_1^2 \xi_2^2q_x^{*4}\boldsymbol{J}_{\text{c}+}$ is much smaller in magnitude, yielding a large value of the SDE. 
For intermediate parameters, such as $\xi_{2}q_x^{*}\sim1$, which corresponds to larger strain, the magnitudes of the two critical currents are comparable, yielding a small value of the SDE. 
Hence, we obtain a field-free SDE along the direction of the applied strain.
Since $\xi_{2}q_{x}^{*}$ is nearly temperature-independent, the resulting diode efficiency is likewise nearly insensitive to temperature~\footnote{We thank A. Levchenko for bringing this aspect of our theory to our attention.}, while it remains strain-dependent.
This is in contrast to the standard SDE case, whose efficiency vanishes as $\sqrt{T_{\text{c}}-T}$~\cite{Daido2022, He2022}.

While these findings are consistent with the observations of Ref.~\cite{Liu2024}, we should critically assess the extent to which they are quantitatively reliable: i) The calculated critical currents are the intrinsic depairing currents and therefore represent upper bounds for homogeneous, defect-free superconductors. In real systems, especially type-II superconductors, dissipation may arise from vortex motion well below the depairing limit, substantially reducing the measured critical current.
ii)  In our analysis, we assumed that the current does not alter the nature of the spontaneous symmetry breaking, i.e., that it does not change the sign of $\bm{q}$.
This assumption is expected to hold in systems whose characteristic dimensions exceed the superconducting penetration depth $\lambda_L$, because the broken symmetry is a bulk effect, whereas the current is confined to a layer of thickness $\lambda_L$.
For materials with dimensions comparable to or smaller than $\lambda_L$, the current-induced nucleation dynamics is likely rather complex and should reduce the diode effect.
In  PbTaSe$_2$, the sample thickness is comparable to $\lambda_L$ while the relevant flake size $\sim10\ \mu\mathrm{m}$ of the material in the lateral direction is significantly larger~\cite{Liu2024}.
Either way, we expect current-induced switching dynamics to be most pronounced near the onset of TRSB, i.e., for $\xi_{2}q_x^{*}\ll1$, where the barrier between time-reversed states is lowest.
This should spoil the perfect diode efficiency, a behavior we would also expect for the systems studied in Refs.~\cite{Yuan2022, Shaffer2024, Chakraborty2025};  see also the discussion in Ref.~\cite{Shaffer2025}.
The barrier height of the helical state is given by $(|a_{1}|^{2}/b)(\xi_{1}/\xi_{2})^{2}(\xi_{2}q_{x}^{*})^{4}$ in the leading order in $q^*$, which is only weakly $T$-dependent.
iii) Close to the critical current, the superconducting order parameter typically varies on the scale of the coherence length.
Consequently, higher-order derivative terms such as $\sim d \psi_1^*\nabla^4\psi_1$ may no longer be negligible.
We added such a higher-order term to our analysis, and Fig.~\ref{Fig_Strain_dep_q_J_SDE} shows the results for different values of the coefficient $d$.
In particular, at larger strains, where $\xi_{2}q_x^{*}\sim1$, the fourth-order derivative term significantly modifies the behavior of the diode effect.
Given these uncertainties, our theory cannot provide quantitative predictions for the magnitude of the SDE. Nevertheless, we expect some level of non-monotonic strain dependence of the SDE to remain observable.
To continue testing our theory, we next analyze the behavior as a function of an external magnetic field and across different strain configurations.

\begin{figure}[h]
    \centering
    \includegraphics[scale=0.65]{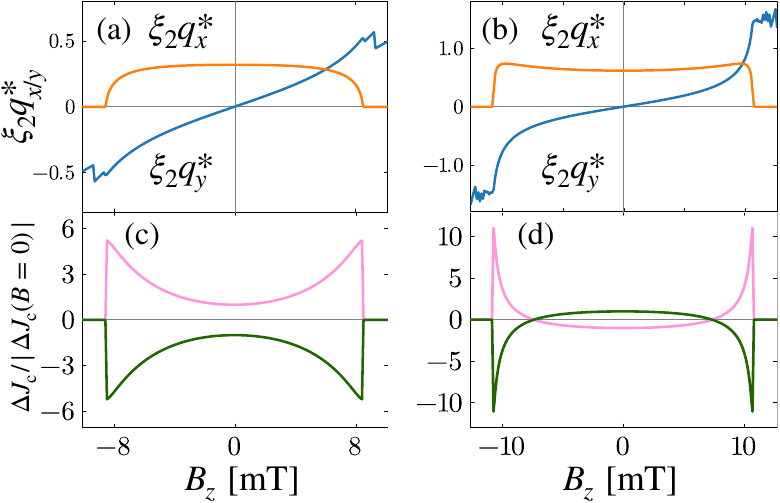}
    \caption{
    (a) and (b): The $B_{z}$-dependence of $\xi_{2}q_{x}^{*}$ (orange) and $\xi_{2}q_{y}^{*}$ (blue).
    (c) and (d): The SDE along the $x$-axis as a function of $B_{z}$ normalized by the absolute value at $B=0$, where the pink (green) curve shows the SDE for TRSB with positive $+q_{x}^{*}$ (negative  $-q_{x}^{*}$) state.
    In panel (a) and (c) $[$(b) and (d)$]$ we chose $\xi_{1}/\xi_{2}=7.07\left[3.16\right]$,
$\lambda/\eta=0.0019\left[0.00083\right]\times\xi_{2}$, $\mu/\eta=3.73\left[1.67\right]\times\xi_{2}$,
and $\xi_{1}^{2}\lambda^{2}/\left(c_{2}c_{1}\right)=1$ and introduce
the characteristic field $B_{0}=\sqrt{|a_{1}|c_{1}}/\nu$, which was
chosen as $B_{0}\approx5\ \mu{\rm T}$ to reproduce the field values observed in Ref.~\cite{Li2025}.
}
    \label{Fig_qxqy_SDE_Bdep}
\end{figure}

\runinhead{Effect of out-of-plane magnetic field}
Applying an out-of-plane magnetic field $B_{z}$ to the system, Ref.~\cite{Liu2024} finds that the SDE along the strain ($x$-axis; armchair direction) exhibits a response that is even in $B_{z}$. This effect occurs for magnetic fields that are smaller than the lower critical field $\mu_0H_{c1}\approx 7.5\,{\rm mT} $~\cite{Ali2014, Sankar2017}, suggesting that it is not an effect dominated by vortex physics.  
For strained systems the additional field-dependent term is allowed:
\begin{equation}
f_B=\mathrm{i}\nu B_{z}\epsilon(\psi_{1}^{*}\partial_{y}\psi_{1} - \psi_{1}\partial_{y}\psi_{1}^{*}).
\end{equation}
An analysis analogous to the one presented above yields
\begin{eqnarray}
    q_{x}^{*}\left(-B_z\right)&=&q_{x}^{*}\left(B_z\right),\nonumber \\
    q_{y}^{*}\left(-B_z\right)&=&-q_{y}^{*}\left(B_z\right).
\end{eqnarray}
The $B_{z}$-dependence of $(q_{x}^{*}, q_{y}^{*})$ for fixed strain is shown in Fig.~\ref{Fig_qxqy_SDE_Bdep} (a) and (b) for different values of $a_{2}$.
While $q_{x}^{*}$ is fairly insensitive to changing $B_{z}$, $q_{y}^*$ is (almost) linear in $B_{z}$. 
This suggests that the field-induced, $B_{z}$-odd helical state develops along the $y$-axis, while the $B_{z}$-even spontaneous helical state remains almost intact, as we increase the strength of $B_{z}$ until $q_{x}^{*}$ disappears.
At fixed $B_z$, TRS is of course broken; a finite value of $q_x^*$ still breaks spontaneously the product of TRS and the mirror reflection $y\rightarrow -y$. 

The $B_{z}$-dependence of the normalized $\Delta J_{\text{c}}$ by the absolute value at $B_z=0$ at fixed strain is shown in Fig.~\ref{Fig_qxqy_SDE_Bdep} (c) and (d) for different $a_{2}$ values, for vanishing fourth derivative term, $d=0$.
Their signs change when we change the sign of the Cooper pair momentum, as expected.
$\Delta J_{\text{c}}$ monotonically develops as the strength of $B_{z}$ increases until $q_{x}^{*}$ vanishes.
The experimentally observed $B_{z}$-dependence~\cite{Liu2024} including a sign change is consistent with Fig.~\ref{Fig_qxqy_SDE_Bdep}(d).
The detailed shape of the curve will again be altered by order-parameter switching dynamics, in particular near the sharp peaks of $\Delta J_{\text{c}}$ where the nucleation barriers are low.

\runinhead{Generic shear strain configurations}
Finally, we consider arbitrary strain configurations $(\epsilon_{1},\epsilon_{2})\equiv \epsilon_{x^{2}-y^{2}},-2\epsilon_{xy})$. 
We consider  $B_{z}=0$ for simplicity. An analysis similar to the one performed earlier yields the  free energy $f(\bm{q}) = -(|a_{1}|^{2}/2b)R^{2}(\bm{q})$, where $R(\bm{q})$ is now 
\begin{equation}
    R(\bm{q}) = -1 + \xi_{1}^{2}q^{2} - \gamma\frac{1 + \tilde{\ell}^{2}q^{2}}{1 + \xi_{2}^{2}q^{2}},
\end{equation}
with $\bm{q}=q(\cos\theta,\sin\theta)$.
We introduce $(\epsilon_{1},\epsilon_{2})=\epsilon(\cos2\phi,-\sin2\phi)$, i.e., $\phi$ is the orientation of the principal strain axis.
The factor $2\phi$ reflects the rank-two transformation of the in-plane strain tensor.
Finally, the angle-dependent length scale is:
\begin{equation}
    \tilde{\ell}^{2}(\theta) = \frac{\mu^{2}}{\eta^{2}}\Big[\big(\Lambda + \cos[2(\theta-\phi)]\big)^{2} + \sin^{2}[2(\theta-\phi)]\Big],
\end{equation}
where  $\Lambda=\lambda/\mu\epsilon$.
The lowest free energy corresponds to the largest value of $\tilde{\ell}$.
For $\Lambda>0$  this implies $\theta=\phi$, while $\theta=\phi+\pi/2$ for $\Lambda<0$.
When $\Lambda \epsilon_1 >0$ and $\epsilon_2=0$, it follows  $\theta=0$, i.e., $\bm{q}=(q,0)$, which is our earlier result.
If we now change the sign of $\epsilon_1=\epsilon_{x^2-y^2}$, i.e., apply compressive rather than tensile strain, we have $\theta=\pi/2$, i.e., $\bm{q}=(0,q)$, which means that the zero-field diode effect should switch by 90 degrees to the $y$ direction.
This is a directly testable prediction. In addition, if one   applies the additional shear strain $\epsilon_{2}=-2\epsilon_{xy}$ the  momentum $\bm{q}$ of the helical state can be rotated continuously according to
\begin{equation}
    \bm{q}= q^{*}\begin{cases}
        (\cos\phi,\sin\phi) & \text{if} \ \Lambda>0 \\
        (-\sin\phi, \cos\phi) & \text{if} \ \Lambda<0
    \end{cases}.
\end{equation}

\runinhead{Summary}
We have developed a symmetry-based Ginzburg--Landau theory of the strain-induced zero-field SDE, with particular emphasis on recent observations in $\textrm{PbTaSe}_{2}$~\cite{Liu2024}.
Strain enhances the Lifshitz coupling of the primary and secondary order parameters and, above a critical threshold, stabilizes a finite-momentum helical state that spontaneously breaks TRS.
This mechanism accounts for both the zero-field SDE and its response to an out-of-plane magnetic field, including the distinct field parities for currents parallel and perpendicular to the mirror plane.

The discontinuous jump to perfect diode efficiency, shown in Fig.~\ref{Fig_Strain_dep_q_J_SDE}, is expected to depend on the nucleation and switching dynamics, and may lead to a smooth but non-monotonic strain variation of the SDE, in particular for sample sizes comparable to the penetration depth.
To test our theory, we propose applying controlled combinations of the two key strain tensor components, $\epsilon_{x^2-y^2}$ and $\epsilon_{xy}$, that rotate the direction in which the SDE can be observed. Compressive strain, in contrast to tensile strain,  reverses this direction to its orthogonal counterpart!

Although formulated for superconductors with point group $D_{3h}$, our framework can be generalized to other point groups by identifying the strain components and order parameters that allow analogous couplings. It thus provides a symmetry-based route for predicting and tuning zero-field SDEs across a broader class of non-centrosymmetric superconductors and, as we showed, could even be obeyed for some materials at zero strain.

\runinhead{Acknowledgment}
We are grateful to Q. Chen, Y. M. Itahashi,  Y. Iwasa,  I. Jang, A. Levchenko, H. Matsuoka, D. Schultz, and Y. Yanase for valuable discussions and comments.
We thank the Wilhelm and Else Heraeus Foundation for organizing and supporting the 855th WE-Heraeus Seminar, {\it Topology and Geometry: Novel Concepts in 3D Quantum Transport on the Mesoscale}, where this work originated in discussions between R.N. and Yoshi Iwasa.
This work was supported by the Toyota Riken Overseas Scholarship (R.N.),
the German Research Foundation TRR 288-422213477 ELASTO-Q-MAT,
 A07 (R.N. and J.S.), and grant  SFI-MPS-NFS-00006741-05 from the Simons Foundation (J.S.).


\bibliography{reference}
\bibliographystyle{apsrev4-2}

\end{document}